\documentclass{article}

\usepackage{microtype}
\usepackage{graphicx}
\usepackage{subfigure}
\usepackage{booktabs} 

\newcommand{\kuruczaone}{\textsl{Kurucz-a1}}

\usepackage{hyperref}

\usepackage[accepted]{icml2025}

\usepackage{amsmath}
\usepackage{amssymb}
\usepackage{mathtools}
\usepackage{amsthm}

\usepackage[capitalize,noabbrev]{cleveref}

\theoremstyle{plain}

\theoremstyle{definition}

\theoremstyle{remark}

\usepackage[textsize=tiny]{todonotes}

\icmltitlerunning{Differentiable Stellar Atmospheres}

\begin{document}

\twocolumn[
\icmltitle{Differentiable Stellar Atmospheres with Physics-Informed Neural Networks}



\icmlsetsymbol{equal}{*}

\begin{icmlauthorlist}
\icmlauthor{Jiadong Li*}{mpia}
\icmlauthor{Mingjie Jian*}{stockholm}
\icmlauthor{Yuan-Sen Ting*}{osu,ccapp}
\icmlauthor{Gregory M. Green}{mpia}
\end{icmlauthorlist}

\icmlaffiliation{mpia}{Max-Planck-Institut für Astronomie, Heidelberg, Germany}
\icmlaffiliation{stockholm}{Department of Astronomy, Stockholm University, Sweden}
\icmlaffiliation{osu}{Department of Astronomy, The Ohio State University, USA}
\icmlaffiliation{ccapp}{Center for Cosmology and AstroParticle Physics, The Ohio State University, USA}

\icmlcorrespondingauthor{Jiadong Li}{jdli@mpia.de}

\icmlkeywords{Machine Learning, Physics-Informed Neural Networks, Stellar Atmospheres, Astrophysics}

\vskip 0.3in
]



\printAffiliationsAndNotice{\icmlEqualContribution} 


\begin{abstract}
We present \textsl{Kurucz-a1}, a physics-informed neural network (PINN) that emulates 1D stellar atmosphere models under Local Thermodynamic Equilibrium (LTE), addressing a critical bottleneck in differentiable stellar spectroscopy. By incorporating hydrostatic equilibrium as a physical constraint during training, \textsl{Kurucz-a1} creates a differentiable atmospheric structure solver that maintains physical consistency while achieving computational efficiency. \textsl{Kurucz-a1} can achieve superior hydrostatic equilibrium and more consistent with the solar observed spectra compared to ATLAS-12 itself, demonstrating the advantages of modern optimization techniques. Combined with modern differentiable radiative transfer codes, this approach enables data-driven optimization of universal physical parameters across diverse stellar populations—a capability essential for next-generation stellar astrophysics.
\end{abstract}

\section{Introduction}\label{sec:intro}

Understanding stellar properties requires detailed modeling of observed spectra using stellar atmosphere models. Since stellar interiors are opaque to radiation, observable absorption features originate primarily from the photosphere \citep{Gray2008}, the layer where light begins to escape from the atmosphere.

Stellar spectral modeling typically involves two central steps: constructing an atmospheric model and generating synthetic spectra. The first step determines the atmospheric structure—temperature, pressure, and electron density as functions of optical depth (measured in Rosseland mean opacity units)—by iteratively solving radiative transfer, radiative and hydrostatic equilibrium equations under assumptions of one-dimensional stratification and local thermodynamic equilibrium (LTE) \citep{Hubeny2015}. 
Traditional approaches rely on pre-computed grids such as ATLAS \citep{Castelli2003}, MARCS \citep{Gustafsson2008}, and PHOENIX \citep{Allard2016}. The second step uses this fixed atmospheric structure for spectral synthesis \citep[e.g.][]{Kurucz1981, Sneden2012, Gerber2023, Piskunov2017},
performing radiative transfer calculations to generate synthetic spectra.

However, the output spectra are influenced by numerous poorly constrained parameters. Atomic transition parameters, such as oscillator strengths for individual lines, often require laboratory measurements or theoretical calculations that may be inaccurate \cite{Asplund_2005, Laverick2018}. Furthermore, atmospheric modeling relies on opacity calculations \citep{Magic2013, Colgan2016}, convection treatments \citep{Asplund2000a, Asplund2000b,Neilson2013,Pasetto2014}, and assumptions about atmospheric structure \citep{Asplund2005},
which introduce systematic uncertainties into spectral modeling.

Large-scale spectroscopic surveys such as the Sloan Digital Sky Survey \citep{York+etal+2000} and LAMOST \citep{Zhao2012} provide extensive datasets that could enable optimization of these model parameters. While individual stars have different fundamental properties (effective temperature, surface gravity, metallicity), the underlying atomic physics remains universal. In principle, a differentiable end-to-end modeling framework could optimize these universal parameters by fitting large stellar samples while marginalizing over star-by-star properties—effectively using the diversity of stellar observations to constrain the physics that governs all stars.

\begin{figure}[ht!]
\centering
\includegraphics[width=0.4\textwidth]{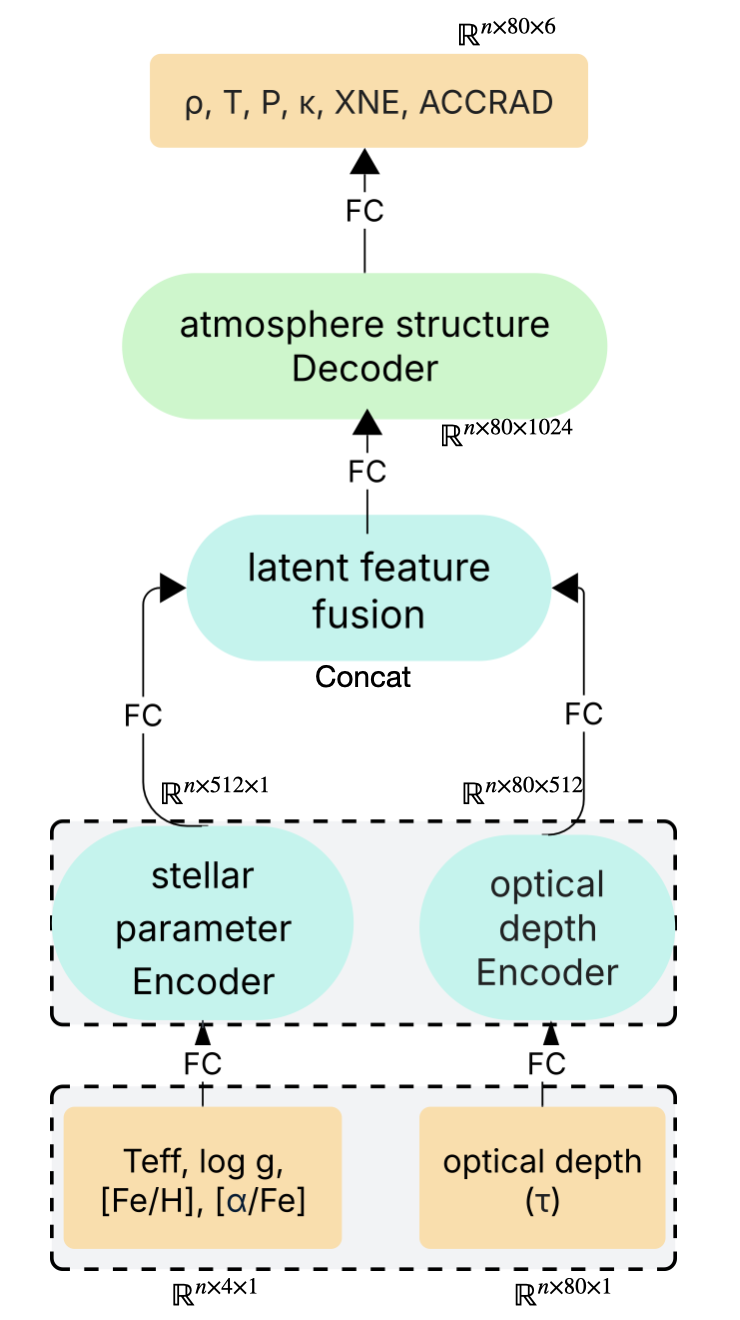}
\caption{\kuruczaone\ neural network architecture with dual-encoder design. Stellar parameters ($T_{\text{eff}}$, $\log g$, [Fe/H], [$\alpha$/Fe]) and optical depth points are encoded separately, then concatenated and processed through fully connected layers (FC) to predict atmospheric parameters ($\rho_x$, $T$, $P$, $\kappa$, $X_{\text{NE}}$, ACCRAD) at each depth point.
}
\vspace{-0.6cm}
\label{fig:architecture}
\end{figure}

Ideally, creating differentiable stellar spectroscopy codes would involve rewriting existing models in modern frameworks like \texttt{PyTorch} or \texttt{JAX}. However, most stellar atmosphere codes predate the widespread adoption of automatic differentiation. Many of these codes \citep[e.g.][]{Castelli2003} are written in legacy languages, such as \texttt{Fortran 77}, and have limited documentation. This presents a barrier to modernization, even with current automated coding tools.

Fortunately, the two modeling components—atmospheric structure model and radiative transfer—can be addressed separately. Modern frameworks such as \texttt{Korg} \citep{Wheeler2024} have successfully modernized radiative transfer solvers in differentiable languages. However, the atmospheric structure solver remains the key bottleneck for end-to-end differentiable modeling.

In this study, we address this bottleneck by developing a physics-informed neural network (PINN) emulator for stellar atmospheric structure calculations. While this approach does not directly solve the differential equations in a differentiable framework, it creates a differentiable component that, when combined with modern radiative transfer solvers, enables the end-to-end optimization framework described above.

\section{Motivation for PINN}\label{sec:motivation}

Emulating stellar atmospheric structure presents challenges even in discretized form. Consider the ATLAS codes \citep{Kurucz1970, Kurucz1981, Kurucz1996, Kurucz2014} —cornerstone tools in stellar astrophysics developed by Robert Kurucz\footnote{Robert Kurucz, who dedicated his life to developing these transformative codes, passed away in 2025.  Together with Fiorella Castelli, they built the foundation on which much of contemporary stellar spectroscopy still relies. This work aims to preserve and further modernize their shared legacy.}. The ATLAS-12 atmospheric structure solver produces high-dimensional vector fields as functions of Rosseland optical depth: six atmospheric parameters (temperature, pressure, density, electron number density, opacity, and radiative acceleration) across 80 depth points, yielding 480-dimensional outputs.

The input space is equally complex. While fundamental stellar parameters ($T_{\text{eff}}$, $\log g$, [Fe/H], [$\alpha$/Fe]) provide the primary constraints, individual elemental abundances can affect atmospheric structure \cite{LeBlanc2009, Ting2016}. This creates a challenging high-dimensional input-output mapping problem where traditional neural network architectures lack appropriate inductive biases for the complex physical relationships governing stellar atmospheres.

\begin{figure*}[h]
\centering
\includegraphics[width=0.99\textwidth]{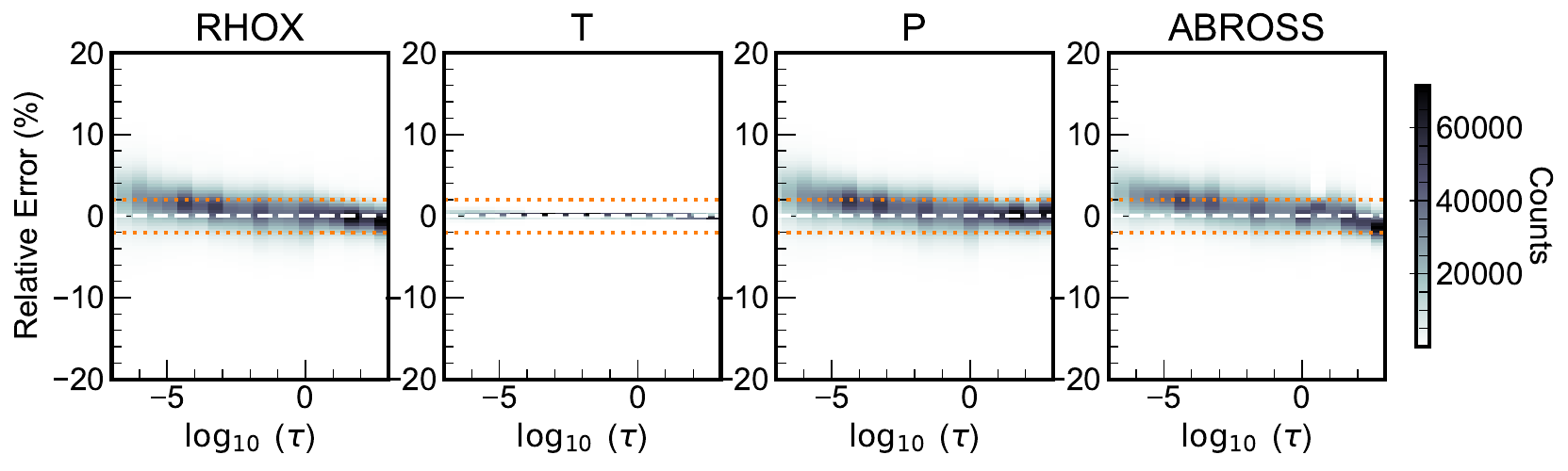}
\caption{Relative error distributions for atmospheric parameters predicted by \kuruczaone compared to ATLAS-12 models. The four panels show column mass density (RHOX), temperature (T), pressure (P), and Rosseland opacity (ABROSS) versus optical depth. The validation dataset spans Galactic stellar populations (see appendix).}
\label{fig:errors}
\end{figure*}

However, stellar atmospheric structure is governed by well-established physical laws—primarily hydrostatic equilibrium and radiative transfer equations. These constraints provide the essential inductive bias that standard architectures cannot capture. Physics-Informed Neural Networks (PINNs) \citep{E2017, Raissi2018, Raissi2019} 
offer an optimal solution by integrating these physical constraints directly into the learning process through composite loss functions that enforce both data fidelity and physics compliance.

\section{Method}\label{sec:method}

\textsl{Kurucz-a1} employs a dual-encoder architecture (Figure~\ref{fig:architecture}) that separates global stellar parameters from local depth information. This design reflects the underlying physics where global stellar properties establish the overall atmospheric structure, while local conditions vary systematically with atmospheric depth.

\begin{figure}[ht!]
    \centering
    \includegraphics[width=0.85\linewidth]{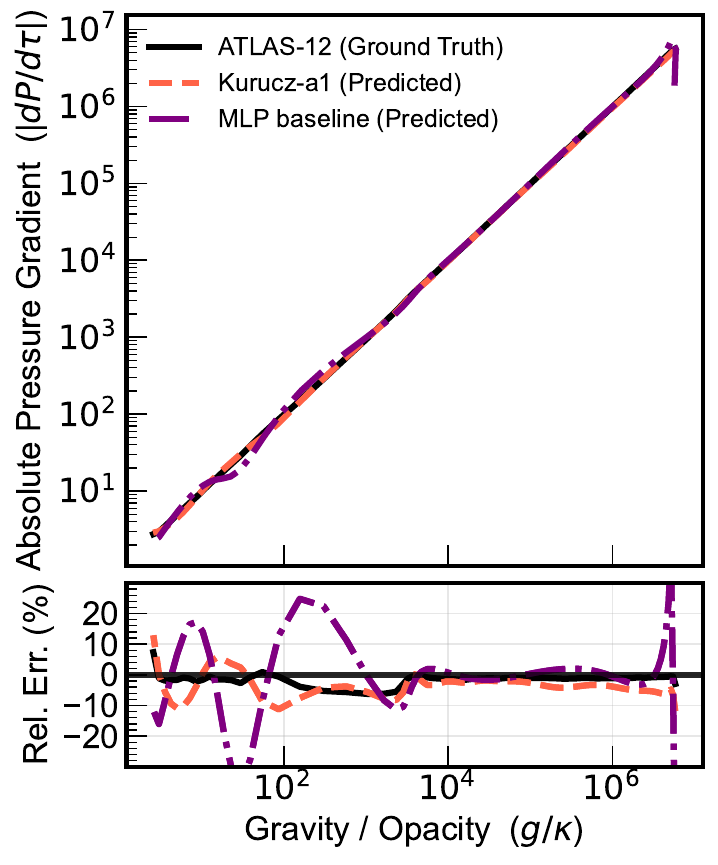}
\caption{Hydrostatic equilibrium diagnostic for a representative red-giant branch star ($T_{\mathrm{eff}} = 4500$\,K, $\log g = 2.5$, $[\mathrm{Fe}/\mathrm{H}] = 0$, [$\alpha$/Fe] = 0). Deviations from the diagonal indicate violations of $dP/d\tau = g/\kappa$. \kuruczaone (dashed) demonstrates superior agreement compared to the MLP baseline (dot-dashed), nearly matching ATLAS-12 (solid).}  
    \label{fig:compare}
\end{figure}

The stellar parameter encoder processes four fundamental quantities—effective temperature ($T_{\text{eff}}$), surface gravity ($\log g$), metallicity ([Fe/H]), and $\alpha$-enhancement ([$\alpha$/Fe])—through a multi-layer perceptron to produce 512-dimensional embeddings. The depth encoder transforms the 80 Rosseland optical depth points ($\tau_{\text{Ross}}$) into 512-dimensional representations at each atmospheric layer, allowing the network to learn appropriate representations for the atmospheric depth coordinate.

The stellar parameter embedding is broadcast and concatenated with each depth embedding to form 1024-dimensional combined embeddings at all 80 depth points. These are processed through a 3-layer MLP emulator (with hidden dimensions [1024, 512, 256]) to predict six atmospheric parameters at each depth: column mass density ($\rho_x$), temperature ($T$), gas pressure ($P$), electron number density ($X_{\text{NE}}$), Rosseland mean opacity ($\kappa_{\text{Ross}}$), and radiative acceleration (ACCRAD). We use GeLU activations \citep{Hendrycks2016}.

\begin{figure}[ht!]
\centering
\includegraphics[width=0.5\textwidth]{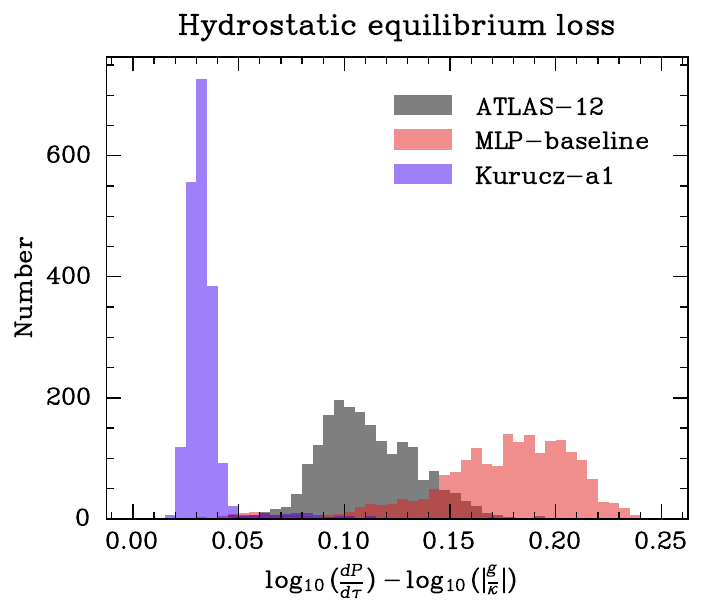}
    \vspace{-0.2cm}
\caption{Distribution of hydrostatic equilibrium loss values across the validation dataset. \kuruczaone (purple) achieves a tight distribution comparable to ATLAS-12 (gray), while the MLP baseline (red) shows higher and more scattered values. Values near zero indicate perfect adherence to the hydrostatic equilibrium constraint.}
   \vspace{-0.5cm}
\label{fig:hydrostatic}
\end{figure}

The total loss function combines data reproduction and physics enforcement:
\begin{equation}
\mathcal{L}_{\text{total}} = (1 - \alpha) \cdot \mathcal{L}_{\text{data}} + \alpha \cdot \mathcal{L}_{\text{physics}}
\end{equation}
with $\alpha = 0.03$ by grid search, where $\mathcal{L}_{\text{data}}$ is the MSE loss of the predicted outputs. The physics loss enforces hydrostatic equilibrium:
\begin{equation}
\mathcal{L}_{\text{physics}} = \frac{1}{N_b \cdot N_d} \sum_{i,j} \left( \frac{dP}{d\tau} - \frac{g}{\kappa} \right)^2_{i,j}
\end{equation}
where $\frac{dP}{d\tau}$ is computed via automatic differentiation, enforcing the constraint that governs atmospheric stability:
\begin{equation}
\frac{dP}{d\tau} = \frac{g}{\kappa}
\end{equation}

\begin{figure*}[ht!]
\centering
\includegraphics[width=\textwidth]{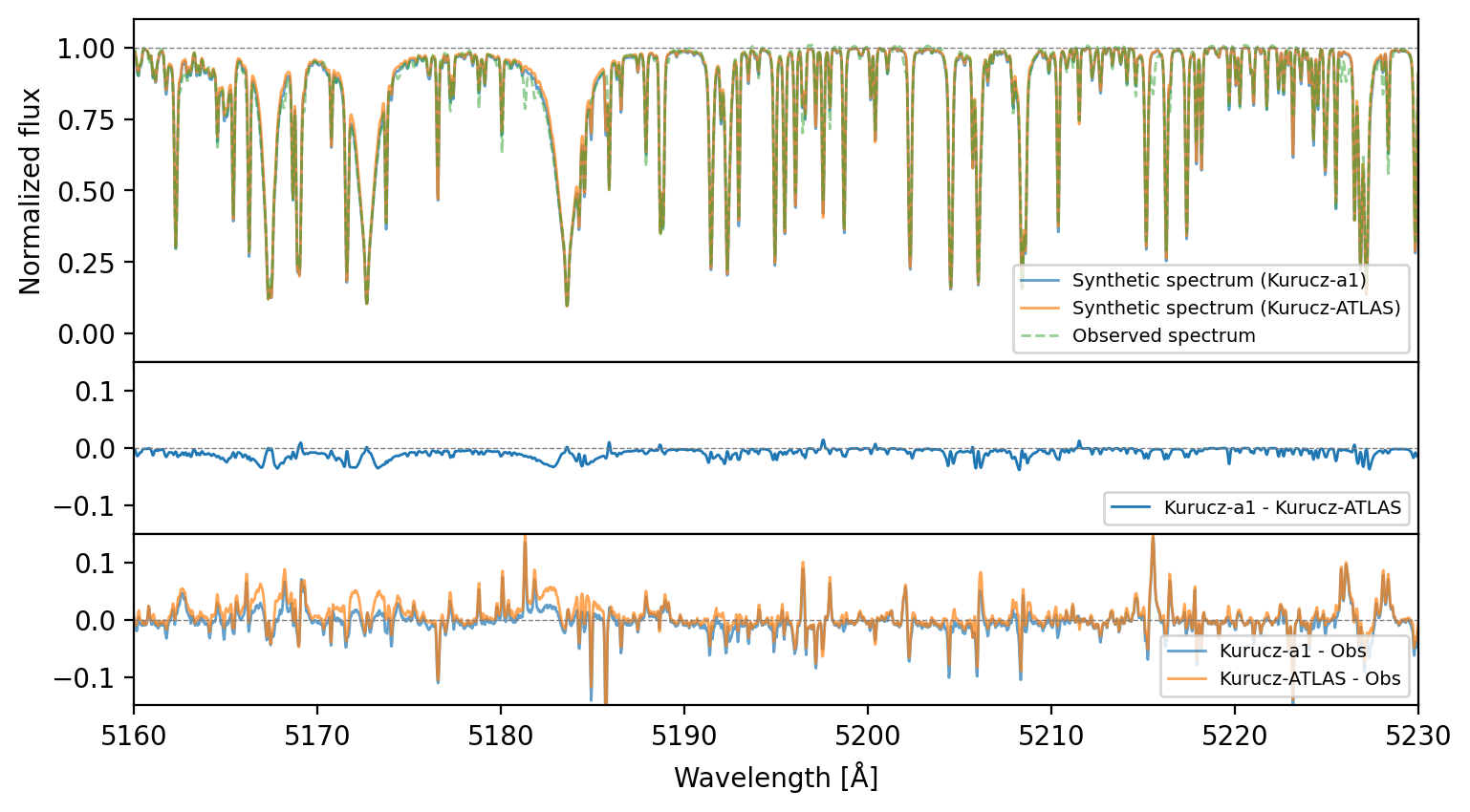}
\vspace{-0.8cm}
\caption{
Solar spectrum synthesis validation. \textbf{Top}: Observed solar spectrum (green dashed) compared to synthetic spectra from \kuruczaone\ (blue) and ATLAS-12 (orange). \textbf{Middle}: Direct comparison between \kuruczaone\ and ATLAS-12 synthetic spectra, showing minimal differences. \textbf{Bottom}: Residuals relative to observed spectrum demonstrate that \kuruczaone\ achieves better agreement than ATLAS-12 due to superior hydrostatic equilibrium adherence.
}
\label{fig:solar-synspec-comparison}
\end{figure*}

\section{Results}

We evaluate \kuruczaone's performance to demonstrate that physics-informed constraints are essential for robust stellar atmosphere emulation. The validation encompasses 10,427 stellar models spanning $T_{\text{eff}} = 2500$--50,000~K, $\log g = -1$--5.5, [Fe/H] $= -4$--$+1.46$, and [$\alpha$/Fe] $= -0.2$--$+0.62$ (see Appendix for training data details).

Figure~\ref{fig:errors} shows the relative error distributions across optical depth for all atmospheric parameters. \kuruczaone achieves median errors below 0.12\% for temperature, 1.1\% for pressure and density, and 1.5\% for opacity across the full range of stellar parameters tested. 
These errors are well within acceptable bounds for stellar atmosphere modeling.
For context, even state-of-the-art grids like MARCS and ATLAS exhibit temperature differences of 10 K to 80 K in their upper layers ($\tau_{\text{Ross}} \le 10^{-2}$), which is considered the current frontier of uncertainty stemming from different input physics \cite{Gustafsson2008}. 
Comparisons for the APOGEE survey show the MARCS and ATLAS9 grids agree to within $<1 \%$ for temperature \cite{Heiter2015} and 2-3\% for pressure \cite{Meszaros2012}.
Since our model's median temperature error is well below these accepted inter-model discrepancies, it demonstrates a high degree of fidelity.

However, accurate stellar atmosphere modeling requires more than point-wise precision—the physical consistency of atmospheric profiles is crucial \citep{Magic2013}. Radiative transfer calculations depend critically on pressure and temperature gradients, not just their absolute values \citep{Hayek2010}. 

Figure~\ref{fig:compare} demonstrates the advantage of \kuruczaone in maintaining hydrostatic equilibrium. \kuruczaone achieves hydrostatic equilibrium comparable to ATLAS-12 itself, while outperforming the MLP baseline that lacks physics constraints. This validates the vital role of physics-informed loss terms.

Notably, \kuruczaone, on average, achieves even better hydrostatic equilibrium than ATLAS-12, likely due to ATLAS-12's discrete numerical scheme and optimization methods that predate modern gradient-based techniques. The small deviations from precise equilibrium in ATLAS-12 ($\sim$0.1\% level) reflect limitations of finite-difference discretization.

Finally, we validated \kuruczaone's practical utility by synthesizing solar spectra using \texttt{PySME} \citep{Wehrhahn2023, Piskunov2017} and comparing with observations (Figure~\ref{fig:solar-synspec-comparison}). The synthetic spectra generated using \kuruczaone closely match both ATLAS-12 models and observations from the Melchior database \citep{Royer2024}, confirming that the emulated atmospheric output is indistinguishable from reference models. \kuruczaone even outperforms ATLAS-12 in some spectral line wings, likely due to better hydrostatic equilibrium adherence.


\section{Broader Impact}

\kuruczaone, available open-source at \url{https://github.com/jiadonglee/kurucz-a1}, provides a practical tool and methodological template for the astronomical community, achieving a mean inference time of $0.37 \pm 0.06$\,millisecond per model on an Apple M1 Pro chip.

For stars hotter than $\sim$4500K, where molecular transitions are minimal, radiative transfer codes typically execute within seconds to minutes. Consequently, stellar spectral analysis commonly employs sparse grids of pre-computed atmospheric models, running only spectral synthesis code like \texttt{Korg} \cite{Wheeler2024} and \texttt{PySME} \citep{Piskunov2017, Wehrhahn2023} 
to synthesize lines for comparison with observations.

However, as demonstrated by \citet{Ting2016}, these sparse atmospheric grids introduce systematic errors in stellar line analysis. While solving stellar atmospheric structure with codes like ATLAS requires computational overhead (minutes to hours per model), current sparse grid approaches also limit analysis to few parameters, grouping elemental abundances into broad categories like [Fe/H] and [$\alpha$/Fe].

\kuruczaone provides a low-cost replacement for stellar atmosphere synthesis models. It generates atmospheric output within seconds, making it practical to eliminate the interpolation errors inherent in sparse grids while maintaining computational efficiency. This framework also eases data storage and transfer requirements when additional abundance dimensions are included.

The current implementation uses four stellar parameters consistent with existing atmospheric grids. Future work will extend the PINN framework to higher-dimensional parameter spaces, enabling incorporation of additional abundance dimensions (alpha-elements, CNO, helium) essential for modeling diverse stellar populations, as well as incorporating additional physics such as equation of state and energy balance.

Beyond practical applications, this work addresses a fundamental bottleneck in achieving end-to-end differentiable stellar spectroscopy. Combined with modern differentiable radiative transfer codes, this approach opens pathways to data-driven calibration of atomic physics parameters that current stellar models must assume.

\section{Acknowledgements}
We thank the anonymous referee for their constructive suggestions.
JL thanks Chao Liu and Hans-Walter Rix for their insightful discussions.
JL acknowledges support from the European Research Council through ERC Advanced Grant No. 101054731.
Computations were performed on the HPC system Raven at the Max Planck Computing and Data Facility.
\bibliography{a1}
\bibliographystyle{icml2025}

\newpage
\appendix
\onecolumn


\section{Appendix: Training Data}

\begin{figure*}
    \centering
    \includegraphics[width=0.9\linewidth]{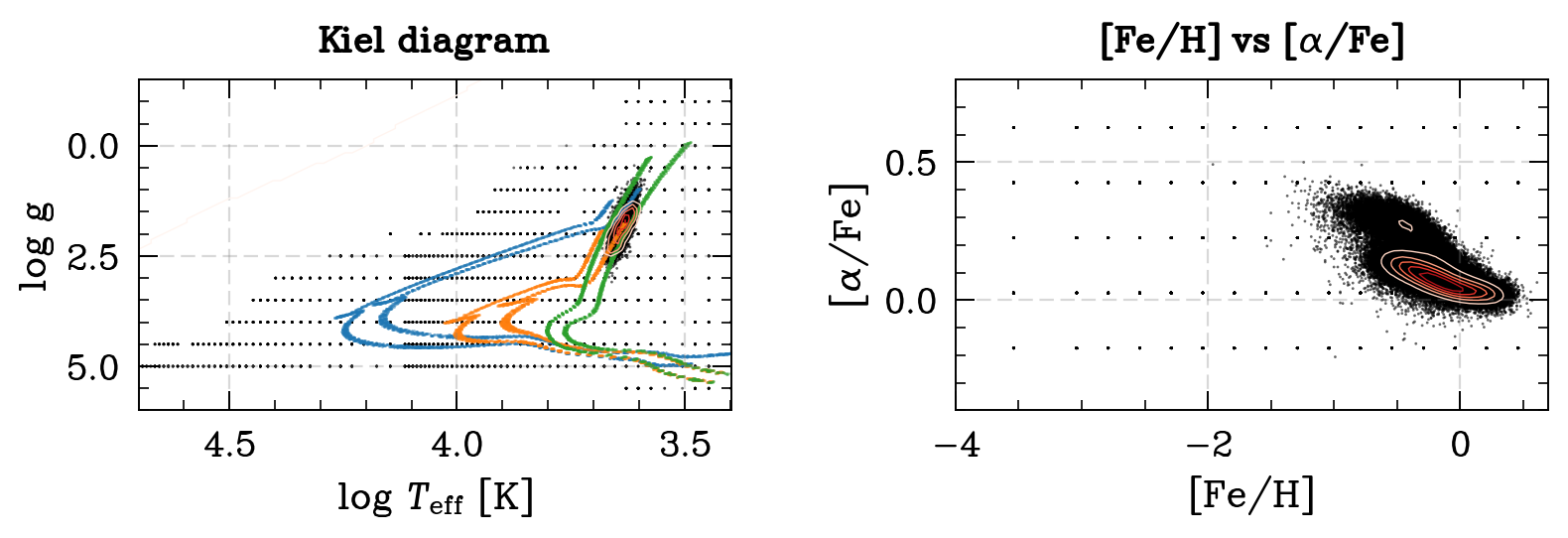}
\caption{Distribution of stellar atmospheric models in the training dataset. \textit{Left panel:} Kiel diagram showing coverage in effective temperature ($\log T_{\text{eff}}$) versus surface gravity ($\log g$) parameter space. The colored evolutionary tracks from PARSEC models \citep{Bressan2012} represent different stellar ages (0.1, 1, 10 Gyr) and metallicities (-1 and 0). \textit{Right panel:} Distribution in the [Fe/H]--[$\alpha$/Fe] abundance plane, showing the density of atmospheric models with contour lines indicating regions of highest concentration.}
    \label{fig:train_data}
\end{figure*}

Our model utilizes pre-computed stellar atmosphere grids from the ATLAS-12 code \citep{2013ascl.soft03024K}, as illustrated in Figure~\ref{fig:train_data}. The grid encompasses effective temperatures ($T_{\text{eff}}$) spanning 2500--50,000~K, surface gravities ($\log g$) from $-1$ to $5.5$, metallicities ([Fe/H]) from $-4$ to $+1.46$, and $\alpha$-element abundances ([$\alpha$/Fe]) from $-0.2$ to $+0.62$. 

The dataset comprises 104,269 atmospheric models, each providing six fundamental parameters as functions of 80 optical depth points: column mass density ($\rho_x$), temperature ($T$), gas pressure ($P$), electron number density ($X_{\text{NE}}$), Rosseland mean opacity ($\kappa_{\text{Ross}}$), and radiative acceleration (ACCRAD).

To optimize sampling for Galactic stellar studies, we implemented enhanced density for red giant branch (RGB) stars in both the Kiel diagram and [Fe/H]--[$\alpha$/Fe] plane, given their importance for understanding Galactic dynamics and chemical evolution. The enhanced sampling around metallicity ([Fe/H] $\sim -1$ to $+0.5$) and $\alpha$-abundance patterns ([$\alpha$/Fe] $\sim 0.0$--$0.4$) reflects the focus on Galactic stellar populations.

The dataset was randomly partitioned into training (90\%, 93,842 models) and validation (10\%, 10,427 models) subsets for model development and performance evaluation.



\end{document}